\documentclass[twocolumn,
preprintnumbers,amsmath,amssymb,prl]{revtex4-1}
\usepackage{amsmath,amssymb,,epsfig}
\usepackage{dcolumn}
\def\be{\begin{equation}}
\def\ee{\end{equation}}
\def\bea{\begin{eqnarray}}
\def\eea{\end{eqnarray}}

\def\half{\frac{1}{2}}

\begin{document}

\preprint{}

\title{The Konishi multiplet at strong coupling}



\author{Brenno Carlini Vallilo}
\email{vallilo@unab.cl}
\affiliation{Departamento de Ciencias F\'\i sicas, Universidad Andres Bello,\\ Republica 220, Santiago, Chile}
\author{Luca Mazzucato}
\email{lmazzucato@scgp.stonybrook.edu}
\affiliation{Simons Center for Geometry and Physics,
Stony Brook University\\
Stony Brook, NY 11794 USA\\}

\date{\today}

\begin{abstract}
We introduce a method to compute from first principles the anomalous dimension of short operators in ${\cal N}=4$ super Yang-Mills theory at strong coupling, where they are described in terms of superstring vertex operators in an anti-de Sitter background. We focus on the Konishi multiplet, dual to the first massive level of the superstring, and compute the one-loop correction to its anomalous dimension at strong coupling, using the pure spinor formalism for the superstring.
\end{abstract}


\maketitle

\section{Introduction}

The computation of the anomalous dimensions of gauge invariant operators is a crucial problem in non-abelian gauge theories. Typically, one can compute such quantities in perturbation theory at weak coupling. Solving this problem in the strong coupling regime or even at finite values of the coupling has been a formidable challenge and requires new insights. In the case of four-dimensional ${\cal N}=4$ super Yang-Mills theory, this seemingly hopeless task can be attacked using the methods of integrability and exploiting the holographic duality between gauge theories and superstring theory \cite{Beisert:2010jr}. 

In the planar limit of the gauge theory, the interplay between these two approaches led eventually to the conjecture that the anomalous dimensions of all gauge invariant operators can be computed at any value of the coupling, using the so called Y-system \cite{Gromov:2009tv}. This conjecture reproduced known results in the sector of long gauge invariant operators, constructed using an asymptotically large number of elementary fields, both at weak and at strong coupling \cite{Beisert:2010jr}. However, the sector of short operators has remained elusive so far.

In this paper, we devise a new method to compute from first principles the anomalous dimensions of operators in ${\cal N}=4$ super Yang-Mills theory at strong coupling. We will describe this new method as applied to short operators, that consist of a small number of elementary fields. The simplest such operator is the $\textrm{Tr}\phi^i\phi^i$, where $\phi^i$ are the six scalars in the super Yang-Mills multiplet. This operator is a member of the long Konishi supermultiplet. 

Short operators at strong coupling are dual to perturbative strings, described in terms of worldsheet vertex operators of type IIB superstring in the $AdS_5\times S^5$ background \cite{Maldacena:1997re,Gubser:1998bc,Witten:1998qj}. Conformal dimensions of field theory operators are equal to energies of such string states. The conformal dimension of non-BPS operators, such as operators in the Konishi multiplet, receive quantum corrections, whose evaluation at strong coupling is the goal of this paper.

By imposing that the worldsheet vertex operators satisfy the superstring physical state condition at the loop level, we derive an equation for the energy of the corresponding string state, which gives in turn the strong coupling expansion of the anomalous dimension of the dual gauge theory operator. We apply our method to the simplest non-BPS operator, a member of the Konishi multiplet, whose anomalous dimension at strong coupling has been predicted in \cite{Gromov:2009tv} using the conjectured Y-system. Our worldsheet computation of the anomalous dimension of particular element gives 
\be
\label{konishi}
\Delta-\Delta_0=2\sqrt[4]{\lambda}-4+{2\over\sqrt[4]{\lambda}}+{\cal O}(1/\sqrt{\lambda}) \ ,
\ee
where $\lambda=g_{YM}^2 N$ is the 't Hooft coupling. The classical dimension $\Delta_0$ of different members of the Konishi multiplet may take different integer values, but the anomalous dimension (\ref{konishi}) should be the same for all of them. Hence, we can pick our favourite member of the multiplet to perform the calculation. The leading term $2\sqrt[4]{\lambda}$ reproduces the expected leading behaviour for an operator dual to a string state at the first massive level \cite{Gubser:1998bc}. The last term $2/\sqrt[4]{\lambda}$ is the one-loop correction to the anomalous dimension of the Konishi multiplet at strong coupling. The three terms in (\ref{konishi}) confirm the numerical fit obtained using the Y-system in \cite{Gromov:2009tv,frolov}, while the first two agree with the bosonic string computation in \cite{Passerini:2010xc} and with the extrapolation from the semiclassical string limit in \cite{Roiban:2009aa}.

Our method can be used to solve for the whole energy spectrum of massive states of type IIB superstring in $AdS_5\times S^5$ and can be expanded to any loop order at strong coupling. This will give an expansion of the anomalous dimensions of short operators in super Yang-Mills theory in inverse powers of $\sqrt[4]{\lambda}$.

\section{String theory in AdS}

The $AdS_5\times S^5$ background is described by the supercoset $PSU(2,2|4)/SO(1,4)\times SO(5)$. A coset representative $g$ transforms as $g\to g_0 g h$ under global $g_0$ and local $h$ transformations. Superstring propagation in this background is described in terms of a worldsheet non-linear sigma model with values in such a supercoset. Since we want to keep covariance manifest, we will make use of the pure spinor action for the superstring in the $AdS$ background (we follow the notations in \cite{Berkovits:2004xu})
\be
\label{adsaction}
S={\sqrt{\lambda}\over 2\pi}\int\, \textrm{Str}[\half J_2\bar J_2+{3\over4}J_3\bar J_1+{1\over4}J_1\bar J_3+
w\bar\nabla l+\bar w\nabla \bar l-N\bar N] \ ,
\ee
where the worldsheet metric is in the conformal gauge. The matter sector of the pure spinor action is constructed in terms of the left-invariant currents $J=g^{-1}d g=\sum_{i=1}^4 J_i$, which take values in the super Lie-algebra $\mathfrak{psu}(2,2|4)$. This algebra admits a $\mathbb{Z}_4$ grading, that we used to label the currents $J_i$. The grading zero part of the current describes the gauge degrees of freedom in the supercoset, namely the local Lorentz rotations, while the even $J_2$ and the odd $J_1,J_3$ describe respectively the ten bosonic and thirty-two fermionic directions of $AdS_5\times S^5$. In addition, the action contains a couple of pure spinor ghosts ${ l},\bar{l}$ of grading one and three respectively, which satisfy the pure spinor constraint $\{{l},{l}\}=0=\{\bar{l},\bar{l}\}$, and their conjugate momenta $w,\bar w$. $N$ and $\bar N$ are the Lorentz generators in the ghost sector. The pure spinor BRST charge is $Q=\oint d\sigma\,\textrm{Str}[{l} J_3+\bar{l}\bar J_1]$ and it is nilpotent on the pure spinor constraint, up to a gauge transformation.  Physical states are in the cohomology of the BRST charge and satisfy the Virasoro constraint $T+\bar T=0$, where $T$ and $\bar T$ are the left- and right- moving worldsheet stress tensors.

The massless sector of type IIB superstring in AdS is described in terms of unintegrated vertex operators of ghost number $(1,1)$ and weight zero, that are in one-to-one correspondence with the type IIB supergravity spectrum in $AdS_5\times S^5$ \cite{Berkovits:2000yr}. In order to study the massive string spectrum, we will expand the sigma model around a classical string configuration, describing a point-like string sitting at the center of AdS. Using the metric of AdS in Lorentzian global coordinates $ds^2=-\cosh^2\rho\,dt^2+d\rho^2+\sinh^2\rho\, dS_3^2$, our string configuration sits at $\rho\sim0$ and evolves in time as $e^{iEt}$. In the static gauge, this is described by the coset element 
\be
\label{ground}
\tilde g(\sigma,\tau)=\exp[-\tau E {\bf T}/ \sqrt{\lambda}] \ ,
\ee
that solves the worldsheet equations of motion coming from the action (\ref{adsaction}), where ${\bf T}$ is the anti-hermitian $PSU(2,2|4)$ generator corresponding to the AdS time translations and $\tau$ is the worldsheet time. The only non-vanishing left-invariant current in this background is $\tilde J_\tau=\tilde g^{-1}\partial_\tau \tilde g=-E{\bf T}/ \sqrt{\lambda}$. Hence, such classical configuration has vanishing BRST charge. 

The Noether charges for the global $PSU(2,2|4)$ symmetry of the string sigma model is given by $Q_{PSU}=\oint d\sigma\, j_\tau$, where 
\be
j_\tau={\sqrt{\lambda}\over 2\pi}g[J_1+J_2+J_3+N+\bar N]_\tau g^{-1} \ .
\ee
In particular, the $AdS$ energy operator $\textsf{E}$ evaluated on the string configuration (\ref{ground}) gives 
\be
\label{energyop}
\textsf{E}=\oint d\sigma\,\textrm{Str} {\bf T}\tilde j_\tau=E \ ,
\ee
where we used $\textrm{Str}({\bf T}{\bf T})=-1$. Since we consider positive energy configurations, we can take $E$ to be positive in the following.

The classical Virasoro constraint for such configuration reads
\be\label{classvir}
T+\bar T={\sqrt{\lambda}\over2}\textrm{Str}\, \tilde J_\tau \tilde J_\tau=-{E^2\over2\sqrt{\lambda}} \ .
\ee
The classical Virasoro constraint (\ref{classvir}) will be modified by quantum effects, which are going to allow for a non-zero solution for $E$ in the rest of the paper. Since for massive string states, such as the one we are considering, the energy scales as $E\sim\sqrt[4]{\lambda}$, the classical contribution to (\ref{classvir}) is of order one and may be canceled against quantum effects.

\section{Quantization}

Let us quantize the action (\ref{adsaction}) around the classical configuration (\ref{ground}) using the background field method. We parameterize the coset element by $g=\tilde g e^X$, where $X=X_1+X_2+X_3$ are the quantum fluctuations and $\tilde g$ is given in (\ref{ground}). We chose a coset gauge in which the grading zero part of the fluctuations vanishes. The left invariant currents are given by
\be
J_\tau= e^{-X}(\partial_\tau-E{\bf T}/ \sqrt{\lambda}) e^X, \qquad J_\sigma=e^{-X}\partial_\sigma e^X \ .
\ee
By expanding the action (\ref{adsaction}) using the $\mathfrak{psu}(2,2|4)$ structure constants up to quadratic order, we can read the spectrum of fluctuations around the background (\ref{ground}). The $AdS_5$ time direction as well as the five sphere directions remain massless, while the remaining four bosonic directions of $AdS_5$ acquire a mass squared  $(E/\sqrt{\lambda})^2$. The fermionic spectrum consists of sixteen massless fermions and sixteen massive fermions with mass squared $(E/2\sqrt{\lambda})^2$. The ghosts remain massless. There is no relation between the bosonic and fermionic spectrum, reflecting the fact that this background is not BPS.

We can canonically quantize the theory imposing the usual equal time commutation relations for the coordinates and their conjugate momenta. If we expand coordinates 
and ghosts into modes following the equations of motion, this will also set the commutation relations between their modes. The vacuum is a scalar and is annihilated 
by all positive modes, including the zero modes of $w,\bar w$. This last requirement ensures that the Lorentz generators for the ghosts $N$ and $\bar N$ annihilate the vacuum.
We can choose sixteen fermionic zero modes as creation operators. Since we are using global AdS coordinates, these are linear combinations of supercharges and superconformal transformations. 
In the rest of the paper, we will evaluate the leading quantum contributions to the physical state condition $T+\bar T$, applied to a particular vertex operator to be introduced below. 

The first quantum correction to (\ref{classvir}) comes from the central charge. Even if for $E=0$, namely in empty AdS, the central charge vanishes \cite{Mazzucato:2009fv}, there is a normal ordering contribution coming from the quadratic part of the stress tensor when $E\neq0$. This is given by the sum of the energies of the oscillator modes
\bea\label{zeropoint}
{2E\over \sqrt{\lambda}}+\half\sum_{n=1}^{\infty}\Bigl(6\sqrt{n^2}+4\sqrt{n^2+(E/\sqrt{\lambda})^2}
\\-16\sqrt{n^2}-16\sqrt{n^2+(E/2\sqrt{\lambda})^2}+22\sqrt{n^2}\Bigr) \ .\nonumber
\eea
The first term is the contribution from the zero modes of the bosons; there is no contribution from the fermionic zero modes, just in the same way as in the pp-wave Hamiltonian \cite{Metsaev:2002re}. Inside the sum, the first two terms come from the bosonic oscillators, the second two terms from the fermionic ones and the last term from the ghosts. We have not computed the precise value of $E$ yet, but we want it to correspond to a stringy state, whose energy scales as $E\sim\sqrt[4]{\lambda}$, which gives $E/\sqrt{\lambda}\sim 1/\sqrt[4]{\lambda}\ll 1$, obtaining from (\ref{zeropoint}) the total contribution $2{E\over\sqrt{\lambda}}-{3\over 16}\zeta(3)({E\over\sqrt{\lambda}})^4$. We can drop the second term as it does not contribute to the energy 
at the order we are considering, so we are left with the contribution 
\be
\label{groundzero}
2{E/ \sqrt{\lambda}} \ .
\ee 
Note that at order $({E\over\sqrt{\lambda}})^2$ the four massive bosonic modes cancel with the massive sixteen fermionic modes. The contribution $2{E\over \sqrt{\lambda}}$, which would vanish in a BPS background, will affect the one-loop correction to the energy of the string, contributing to the last term in (\ref{konishi}).

Let us consider now a specific worldsheet vertex operator. All of the members of the Konishi multiplet have the same anomalous dimension and they are in one to one correspondence with the string states at the first massive level \cite{Roiban:2009aa}. Thus we will choose a particularly simple state in the first massive string level, that will simplify the computation. Physical states are given by unintegrated vertex operators of ghost number $(1,1)$. The simplest one is $\textrm{Str}\,l \bar l$ and it corresponds to the radius modulus at zero momentum \cite{nathantop}. We will denote the corresponding  state as $|l\bar l\rangle\equiv\textrm{Str}(l\bar l)|0\rangle$. 

We choose the simple state
\be\label{vertex}
|V\rangle=x_{-1}^+\bar x_{-1}^+|l\bar l\rangle \ ,
\ee
where $x_{-1}^+$ and $\bar x_{-1}^+$ are the first left- and right-moving oscillators coming from the fluctuations of the $AdS$ ``space-cone'' coordinate $x^+=x^1+ix^2$. Although it does not look covariant, we can interpret this state as being created by non-zero modes of the global symmetry right invariant currents. We should emphasize that this is {\em not} a global $PSU(2,2|4)$ transformation. We identify the vertex operator (\ref{vertex}) to be dual to a particular member of the  Konishi multiplet with classical dimension $\Delta_0=6$, Lorentz spin two and singlet of $SU(4)$. The operator $|l\bar l\rangle$, corresponding to the radius changing operator, is dual to the Yang-Mills lagrangian and it has $\Delta_0=4$, so it is natural to expect that (\ref{vertex}) has two unites more of classical dimension. The same type of state was discussed in \cite{nathantop}, where it is argued that such states are physical, {\em i.e.} they are annihilated by the BRST charge. This state is a two-magnon impurity of mass $E/\sqrt{\lambda}$, whose contribution to the worldsheet energy is 
\be
\label{magnon}
2\sqrt{1+({E/\sqrt{\lambda}})^2} \ .
\ee

\section{Quartic corrections}

Other possible contributions to the physical state condition at this order may come from the terms in the stress tensor $T+\bar T$, expanded to quartic order in fluctuations around the classical background (\ref{ground}) and acting on the specific vertex operator (\ref{vertex}). Let us analyze the possible terms.

The factor of $2$ in front of the square root in (\ref{magnon}) may get corrected by quartic terms in $T+\bar T$ of the form $(\partial X)^2 X^2$, due to normal ordering. However, there is no normal ordering due to this term. This comes from the fact that any correction to this term has to be proportional to the one-loop beta function, which vanishes \cite{Vallilo:2002mh}. The same type of normal ordering 
contribution was discussed in \cite{Callan:2003xr}, where they argued it should be zero using $PSU(2,2|4)$ symmetry 
(the same reason why the beta function vanishes). There are other corrections that are not protected by the beta function argument, but they are of the form $(E/\sqrt{\lambda})\partial X X^3$ and $(E/\sqrt{\lambda})^2 X^4$. They give higher order contributions to the energy and we can safely neglect them.

The last possible contribution comes from the fact that the operator (\ref{vertex}) might mix with other operators due to quartic terms in $T+\bar T$. In order to study the mixing, we have to compute the momenta conjugate to the fields up to   quartic terms in the action, then plug these back in the stress tensor \cite{Callan:2003xr}. The conjugate momenta are given by
\be
P_i={\delta S\over \delta \partial_\tau X_i}=\partial_\tau X_i+\ldots \ 
\ee
where '$\ldots$' are higher order terms in the fluctuations $X_i$. In this way we eliminate all the time derivatives in the stress tensor. For the particular vertex operator (\ref{vertex}) we may only consider the terms with four bosonic or two bosonic and two fermionic fields.
Terms quartic in bosons can both introduce mixing and also correct the energy of our state. Terms with 
two fermions and two bosons will only give mixing. For the particular choice of the ``space-cone'' polarization in (\ref{vertex}), it is easy to see that there will be no mixing with other bosonic states, nor mixing with states created by two fermions, since the stress tensor will only have commutators and products of gamma matrices, which vanish for this choice of polarization. However, there is a non-vanishing correction to the energy, proportional to the state itself, coming from the term ${1\over6}(\textrm{Str}\,[P_2,X_2]^2-\textrm{Str}\,[\partial_\sigma X_2, X_2]^2   )$ in the stress tensor. Expanding 
the Hamiltonian into modes and computing the relevant terms one finds that the corresponding correction is (it coincides with the result in \cite{Callan:2003xr})
\be
\label{quartic}
-2/\sqrt{\lambda} \ .
\ee

\section{Conformal dimension}

Summing up the contributions (\ref{classvir}), (\ref{groundzero}), (\ref{magnon}), and (\ref{quartic}) to the Virasoro constraint, we find that the physical state condition
\be
(T+\bar T) |V\rangle =0 \ ,
\ee
gives
\be
\label{marginality}
-{E(E-4)\over2\sqrt{\lambda}}+2\sqrt{1+({E\over\sqrt{\lambda}})^2}-{2\over\sqrt{\lambda}} =0 \ .
\ee
The positive energy solution of this equation gives the energy of our string state (\ref{vertex}). According to the AdS/CFT dictionary, the energy operator on the string side of the correspondence is mapped to the dilatation operator on the field theory side, whose eigenvalues are the conformal dimensions of operators. Above, we identified the field theory dual to the string state (\ref{vertex}) as a member of the Konishi multiplet with classical dimension $\Delta_0=6$, Lorentz spin two and singlet of $SU(4)$. Its conformal dimension at strong coupling is therefore
\be 
\Delta =E= 2\sqrt[4]{\lambda}+2+{2\over\sqrt[4]{\lambda}}+{\cal O}(\lambda^{-1/2}) .
\ee
Hence we derived (\ref{konishi}) with $\Delta_0=6$.

\noindent{\bf Acknowledgments}:~
We would like to thank N.~Berkovits, O. Chand\'\i a, N.~Gromov, V.~Kazakov, W. Linch, A.~Pakman,  L.~Rastelli, S.~Razamat, R.~Roiban, A.~Tseytlin and B.~van Rees for discussions. We are especially grateful to J.~Maldacena for suggesting the expansion around a classical configuration and to P.~Vieira for invaluable help with Mathematica. We also thank J.~Maldacena, R.~Roiban, A.~Tseytlin and P.~Vieira for comments on the manuscript and N.~Gromov for making the results of \cite{kolya} available to us prior to publication. BCV would like to thank the Simons Center of Geometry and Physics at Stony Brook University for the hospitality at various stages of this project. The authors would like to thank the KITP in Santa Barbara for the hospitality at the 2009 Workshop on Fundamental Aspects of Superstring Theory, where this project was initiated. LM is supported in part by DOE grant DE-FG02-92ER40697.

\noindent {\bf Note Added:}~
After the completion of this manuscript, we became aware that the authors of \cite{kolya} and \cite{roiban} computed the one-loop correction to the anomalous dimension of the Konishi operator at strong coupling using different methods, obtaining the same results as in this paper.


\end{document}